\documentclass{aa}  
\usepackage[varg]{txfonts}
\usepackage{algorithm}
\usepackage{algpseudocode}
\usepackage{flushend}

\begin{document} 

\title{Space variant deconvolution of galaxy survey images}

\author{S. Farrens \thanks{email: samuel.farrens@cea.fr}
		\and F.M. Ngol\`{e} Mboula 
		\and J.-L. Starck}

\institute{Laboratoire AIM, UMR CEA-CNRS-Paris 7, Irfu, Service d'Astrophysique, CEA Saclay, F-91191 GIF-SUR-YVETTE Cedex, France} 
	 
\date{}

\abstract{Removing the aberrations introduced by the Point Spread Function (PSF) is a fundamental aspect of astronomical image processing. The presence of noise in observed images makes deconvolution a nontrivial task that necessitates the use of regularisation. This task is particularly difficult when the PSF varies spatially as is the case for the Euclid telescope. New surveys will provide images containing thousand of galaxies and the deconvolution regularisation problem can be considered from a completely new perspective. In fact, one can assume that galaxies belong to a low-rank dimensional space. This work introduces the use of the low-rank matrix approximation as a regularisation prior for galaxy image deconvolution and compares its performance with a standard sparse regularisation technique. This new approach leads to a natural way to handle a space variant PSF. Deconvolution is performed using a Python code that implements a primal-dual splitting algorithm. The data set considered is a sample of 10\,000 space-based galaxy images convolved with a known spatially varying Euclid-like PSF and including various levels of Gaussian additive noise. Performance is assessed by examining the deconvolved galaxy image pixels and shapes. The results demonstrate that for small samples of galaxies sparsity performs better in terms of pixel and shape recovery, while for larger samples of galaxies it is possible to obtain more accurate estimates of the galaxy shapes using the low-rank approximation.}

\keywords{Methods: numerical -- Techniques: image processing -- Surveys}

\maketitle

%-------------------------------------------------------------------

 \section{Introduction}
 \label{sec:intro}

Deconvolution has been an indispensable mathematical tool in signal and image processing for several decades. In diverse fields such as medical imaging and astronomy accurate and unbiased knowledge of true image properties is paramount. All optical systems, however, are subject to imperfections that distort the images. The sum of these aberrations is commonly referred to as the Point Spread Function (PSF).

Many methods have been devised over the years to deconvolve the effects of a known PSF from an observed image. The most popular algorithm in astrophysics is certainly the Richardson-Lucy algorithm \citep{richardson:1972, lucy:1974}, which iteratively searches for the maximum likelihood solution assuming Poisson noise. This algorithm has been applied to a variety of different topics such as the dust trails of comets \citep{sykes:91}, the inner properties of M31 \citep{kormendy:99}, the X-Ray remnants of supernovae \citep{burrows:00}, the mass distribution of exoplanets \citep{jorissen:01}, the shape estimation of galaxies for weak lensing analysis \citep{kitching:08} and the primordial power spectrum \citep{hamann:10}. The major drawback of the Richardson-Lucy algorithm is that it is not regularised and it converges (slowly) to a noisy solution. In practice, the user generally stops the algorithm before convergence, after a arbitrary number of iterations. This can be considered as a form of regularisation, but it is not very efficient \citep{starck:book06}. Another example is the CLEAN algorithm \citep{hogbom:74}, which assumes objects are comprised of point sources and has been applied to the study of extragalactic radio sources \citep{miley:80}. Early implementations of regularisation in deconvolution problems are the Maximum Entropy Method (MEM) of \citet{gull:78} and the Pixon method \citep{dixon:96}, which models objects as the sum of pseudoimages but has the tendency to overregularise fainter sources. Sparsity has emerged as a extremely powerful approach to regularise inverse problems in general, including deconvolution, especially when using wavelets for representing the data \citep{starck:book15}. For example, it has been shown with LOFAR data that sparsity improves the resolution of restored images by a factor of two compared to the standard  CLEAN algorithm \citep{starck:garsden2015}. Similarly, introducing wavelets into the Richardson-Lucy algorithm \citep{starck:sta94_1,starck:mur95_2} or into the Maximum Entropy Method \citep{starck:sta99_2} has been shown to be extremely efficient. See \citet{starck:02} for an in-depth review of various deconvolution techniques and their applications to astronomical data.

Very few efficient methods have been proposed so far for the case in which the PSF is varying spatially. The simplest approach consists in partitioning the image into overlapping patches and then independently deconvolving each patch with the PSF corresponding to its centre. A more elegant approach would consist in having an {\em Object-Oriented Deconvolution} \citep{starck:sta00_3}. In this case, the assumption is made that the objects of interest can first be detected using software like SExtractor \citep{bertin:96} and then each object can be independently deconvolved using the PSF associated to its centre. In this paper, this concept is extended and, using the premise that galaxies belong to a low dimensional manifold, it is shown that an object-oriented deconvolution also leads to a new way to regularise the problem.

This work additionally investigates the novel idea of low-rank galaxy penalisation and compares the results with the state-of art deconvolution algorithm, namely sparsity, that employs sparse wavelet prior knowledge to aide in the deconvolution of galaxy survey images. The data used for these tests consist of a catalogue of space-based galaxy images convolved with a Euclid-like spatially varying PSF. 

This paper is organised as follows. Section~\ref{sec:bg} provides a brief introduction to some of the mathematical techniques commonly implemented in space variant deconvolution. Section~\ref{sec:lowr} introduces the concept of using the low-rank approximation to regularise a deconvolution problem. Section~\ref{sec:image_rec} describes the optimisation algorithm used to implement the deconvolution techniques. The data used to test these techniques is described in Sect.~\ref{sec:data} and Sect.~\ref{sec:data_app} shows the results of the application to the data. Finally, Sect.~\ref{sec:conc} presents the overall conclusions taken from this work.

%--------------------------------------------------------------------

\section*{Notation}
\label{sec:not}

The following notation conventions are adopted throughout this paper:
	
\begin{itemize}
    \item bold lower case letters are used to represent vectors;
    \item bold capital case letters are used to represent matrices;
    \item vectors are treated as column vectors unless explicitly mentioned otherwise. 
\end{itemize}

$\mathcal{L}^*$ denotes the adjoint operator of a linear operator $\mathcal{L}$. $\rho(\cdot)$ denotes the spectral norm (\emph{i.e.} the largest singular value) of a matrix or a linear operator.  

The $i^{th}$ coefficient of a vector $\mathbf{x}$ is denoted by $\mathbf{x}_i$. The coefficient $(i,j)$ of a matrix $\mathbf{M}$ is denoted by $\mathbf{M}_{i,j}$. $\mathbf{M}_{:,j}$ and $\mathbf{M}_{i,:}$ represent the $j^{th}$ column and $i^{th}$ row of $\mathbf{M}$, which are treated as column and row vectors respectively.

The underlying images are written in lexicographic order (\emph{i.e.} lines after lines) as column vectors of pixels values and the $2D$ convolution operations are represented by matrix-vector products. Each image is comprised of $p$ pixels.

%--------------------------------------------------------------------

\section{Space variant deconvolution}
\label{sec:bg}

\subsection{Linear inverse problem}

The process of deconvolving an observed image that contains random noise and for which the PSF of the optical system is known is equivalent to solving the linear inverse problem 

\begin{equation}
	\mathbf{y} = \mathbf{H}\mathbf{x} + \mathbf{n},
	\label{eq:lip}
\end{equation}

\noindent where $\mathbf{y}$ is the observed noisy image, $\mathbf{x}$ is the signal (\emph{i.e.} the ``true'' image) to be recovered, $\mathbf{n}$ is the noise content and $\mathbf{H}$ represents the convolution with the PSF.

For the purposes of this work, galaxy images are assumed to be those that can be detected in a typical galaxy survey using source extraction software such as SExtractor \citep{bertin:96}. In other words, the intergalactic medium is neglected. 

For a survey of galaxy images let $(\mathbf{y}^i)_{0\leq i\leq n}$ denote the set of detected galaxies and $(\mathbf{x}^i)_{0\leq i\leq n}$ the corresponding true galaxy images. Eq.~\ref{eq:lip} can then be reformatted for the case in which the PSF varies as a function of position on the sky (hereafter referred to as a \emph{space variant PSF}) as

\begin{equation}
	\mathbf{Y} = \mathcal{H}(\mathbf{X}) + \mathbf{N},
	\label{eq:lip2}
\end{equation}

\noindent where $\mathbf{Y} = [\mathbf{y}^0, \mathbf{y}^1, \dotsc, \mathbf{y}^n]$, $\mathbf{X} = [\mathbf{x}^0, \mathbf{x}^1, \dotsc, \mathbf{x}^n]$, $\mathbf{N} = [\mathbf{n}^0, \mathbf{n}^1, \dotsc, \mathbf{n}^n]$ is the noise corresponding to each image and $\mathcal{H}(\mathbf{X}) = [\mathbf{H}^0\mathbf{x}^0, \mathbf{H}^1\mathbf{x}^1, \dotsc, \mathbf{H}^n\mathbf{x}^n]$ is an operator that represents the convolution of each galaxy image with the corresponding PSF for its position.

In order to solve a problem of this type one typically attempts to minimise some convex function such as the least squares minimisation problem

\begin{equation}
    \begin{aligned}
        & \underset{\mathbf{X}}{\text{argmin}}
        & \frac{1}{2}\|\mathbf{Y}-\mathcal{H}(\mathbf{X})\|_2^2,
    \end{aligned}
    \label{eq:l2min}
\end{equation}

\noindent which aims to find the solution $\hat{\mathbf{X}}$ that gives the lowest possible residual ($\mathbf{Y} - \mathcal{H}(\hat{\mathbf{X}})$).

This problem is ill-posed as even the tiniest amount of noise will have a large impact on the result of the operation. Therefore, to obtain a stable and unique solution to Eq.~\ref{eq:lip2}, it is necessary to regularise the problem by adding additional prior knowledge of the true images.
            
\subsection{Positivity prior}

A very simple constraint that can be imposed upon Eq.~\ref{eq:l2min} is that all of the pixels in the reconstructed images have positive values which gives    

\begin{equation}
    \begin{aligned}
        & \underset{\mathbf{X}}{\text{argmin}}
        & \frac{1}{2}\|\mathbf{Y}-\mathcal{H}(\mathbf{X})\|_2^2
        & & \text{s.t.}
        & & \mathbf{X} \ge 0,
    \end{aligned}
    \label{eq:l2minpos}
\end{equation}    

\noindent where the inequality is entry-wise. This is a very sensible assumption as it is known a priori that the true images cannot have negative pixel values.

This prior is used for all the minimisation problems implemented in this paper and is hereafter referred to as a \emph{positivity constraint}.

\subsection{Sparsity prior}    

A powerful regularisation constraint that can be applied to a large variety of inverse problems is the prior knowledge that a given signal can be sparsely represented in a given domain. In other words, if it is known that the signal that one aims to recover is sparse (\emph{i.e.} most of the coefficients are zero) when acted on by a given transform (\emph{e.g.} Fourier, wavelet, \emph{etc.}), one can impose that the recovered signal must also be sparse under the same transformation. This significantly limits the possible solutions of the minimisation problem. 

The exact sparsity of a signal can be measured with the $l_0$ pseudo-norm ($\| \cdot \|_0$), which simply counts the number of non-zero elements in the signal. This, however, is computationally hard to solve in practical applications (specifically NP-hard) and thus sparse solutions are generally promoted via the $l_1$ norm,
    
\begin{equation}
	\|\mathbf{x}\|_1 =  \sum_{i=0}^{p-1} |\mathbf{x}_i|.
\end{equation}

A clear link can be seen with the idea behind the MEM method. Indeed, here the amount of information contained in a solution is measured by the $l_1$ norm. To solve the problem a solution must be found that is both compatible with the data and that contains the smallest amount of information possible. A hugely compelling argument for sparse regularisation is the compressive sensing theorem. This theorem demonstrates that, under certain conditions regarding the signal $\mathbf{x}$ and the operator $\mathbf{H}$, a perfect reconstruction can be achieved through  $l_1$ minimisation. Such a theorem does not exist for any other regularisation technique. In most cases the required conditions are not satisfied, but one can still see the compressed sensing theorem as an asymptotical behaviour of the problem.

Galaxy images are not sparse in the pixel domain. It is therefore useful to introduce a matrix $\boldsymbol{\Phi}$ that transforms the images into a domain which is more sparse. So, for a galaxy image represented by a vector $\mathbf{x}$, the vector $\boldsymbol{\Phi}\mathbf{x}$ is approximately sparse.

Adding the corresponding $l_1$ constraint to Eq.~\ref{eq:l2minpos} gives the minimisation problem

\begin{equation}
    \begin{aligned}
        & \underset{\mathbf{X}}{\text{argmin}}
        & \frac{1}{2}\|\mathbf{Y}-\mathcal{H}(\mathbf{X})\|_2^2 + \lambda\|\Phi(\mathbf{X})\|_1
        & & \text{s.t.}
        & & \mathbf{X} \ge 0
    \end{aligned}
\label{eq:sparsel2minpos}
\end{equation}   

\noindent where $\Phi(\mathbf{X}) = [\boldsymbol{\Phi}\mathbf{x}^0, \boldsymbol{\Phi}\mathbf{x}^1,\dotsc,\boldsymbol{\Phi}\mathbf{x}^n]$, $\|\Phi(\mathbf{X})\|_1 \overset{\text{def}}{=} \sum_{i=0}^n \|\boldsymbol{\Phi}\mathbf{x}^i\|_1$ and $\lambda$ is a regularisation control parameter.

For complex signals, such as images, wavelets provide an excellent basis for sparse decomposition. This is because wavelets provide simultaneous information about the scale and position of features in a given image which in turn leads to better control of the noise.    

Sparse regularisation using wavelets has been very successful when applied to topics such as weak lensing mass mapping \citep{leonard:14, lanusse:16}, the analysis of the cosmic microwave background \citep{bobin:14} and PSF reconstruction \citep{ngole:14}. See \citet{starck:book15} and references therein for a comprehensive review of sparse regularisation. 

\subsection{Sparsity implementation}

\begin{figure}
    \centering
    \includegraphics[width=8cm]{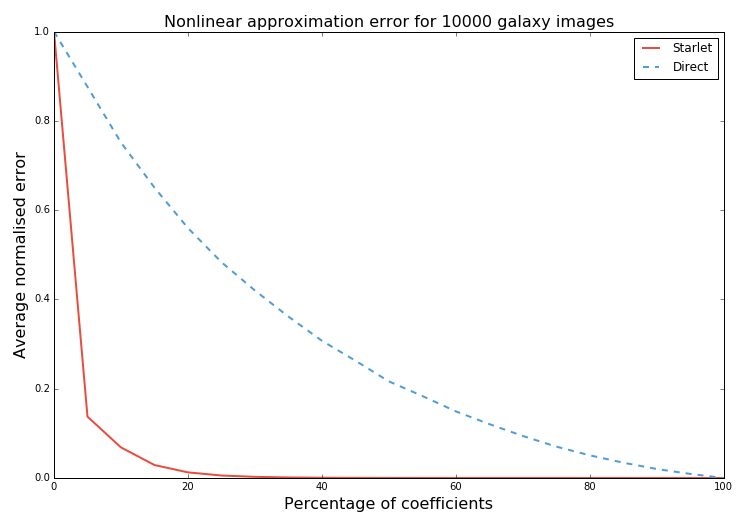}
    \caption{The average nonlinear approximation error as function of percentage of largest coefficients for 10\;000 galaxy images. The solid red line denotes the decay in the starlet space and the blue dashed line denotes the decay in the direct space.}
    \label{fig:nla}
\end{figure}

For this work, sparse representations of the galaxy images are obtained using the  starlet transform \citep[\emph{i.e} an isotropic undecimated wavelet transform,][]{starck:15}. The efficiency of this sparse decomposition is demonstrated in fig.~\ref{fig:nla}, which shows the average nonlinear approximation error for the galaxy images as a function of the percentage of largest coefficients \citep[see][chap.~8]{starck:book15}. The solid red line denotes the decay in the starlet space and the blue dashed line denotes the decay in the direct space. The figure clearly shows a faster decay for the starlet decomposition.

The starlet transform decomposes an image, $\mathbf{x}$, into a coarse scale, $\mathbf{x}^J$, and wavelet scales, $(\boldsymbol{w}^j)_{1\leq j \leq J}$,

\begin{equation}
    \mathbf{x} = \mathbf{x}^{J} + \sum_{j=1}^J \boldsymbol{w}^{j},
\end{equation}

\noindent where the first level ($j = 1$) corresponds to the highest frequencies (\emph{i.e.} the finest scale). The starlet transform is well suited to most astronomical images which are generally isotropic. The starlet transformation implemented in this work is that provided in the iSAP\footnote{http://www.cosmostat.org/software/isap/} software package using $J=3$ and neglecting the coarse scale.

Minimisation with sparse regularisation is implemented by solving a sequence of problems of the form

\begin{equation}
    \begin{aligned}
        & \underset{\mathbf{X}}{\text{argmin}}
        & \frac{1}{2}\|\mathbf{Y}-\mathcal{H}(\mathbf{X})\|_2^2 + 
        \|\mathbf{W}^{(k)}\odot\Phi(\mathbf{X})\|_1
        & & \text{s.t.}
        & & \mathbf{X} \ge 0
    \end{aligned}
    \label{eq:rw_l1}
\end{equation}

\noindent where $\Phi$ realises the starlet transform without the coarse scale, $\mathbf{W}^{(k)}$ is a weighting matrix, with $0 \leq k \leq k_{\max}$ and $\odot$ denotes the Hadamard (entrywise) product. Thus, $\mathbf{W}^{(k)}$ and $\Phi(\mathbf{X})$ are both $J*p \times n+1$ matrices. $k$ is a reweighting index.
 
The $l_1$ norm is generally implemented via soft-thresholding (see Eq.~\ref{eq:soft}). This means that the smallest coefficients will be set to zero and the largest coefficients, which contain the most useful information, will have reduced amplitudes. This can lead to a biased reconstruction of the original signal. Therefore, to alleviate this imbalance and more closely approximate the $l_0$ norm, the reweighting scheme described in \citet{candes:07} is implemented. Let $\hat{\mathbf{X}}^{k}$ denote the solution to Eq.~\ref{eq:rw_l1} for a given $k$. The weights are defined by the recurrence relation

\begin{equation}
    \mathbf{W}_{i,j}^{(k+1)} = \mathbf{W}_{i,j}^{(k)} \frac{1}{1 + \frac{|\Phi(\hat{\mathbf{X}}^{(k)})_{i,j}|}{\mathbf{W}_{i,j}^{(0)}}}
\end{equation}

\noindent where $\mathbf{W}^{(0)}$ denotes the initial weighting matrix. $\mathbf{W}^{(0)}$ is set according to the uncertainty that propagates to the estimated galaxies wavelet coefficients in the deconvolution process. The idea is to assign strong weights to the wavelet coefficients that are severely affected by the observational noise and weaker weights to the others. To do so, the wavelet matrix is written as

\begin{equation}
    \boldsymbol{\Phi} = [\mathbf{\Phi}^{1 T},\cdots,\mathbf{\Phi}^{J T}]^T,
    \label{eq:wav_mat}
\end{equation}	    
	
\noindent where $\mathbf{\Phi}^{j}\mathbf{x}^i$ is the $j^{th}$ (undecimated) wavelet scale of the image $\mathbf{x}^i$.

If the PSF convolution matrices, $\mathbf{H}^{i}$, were invertible and well-conditioned, a straightforward restoration procedure would consist in denoising the vector $\mathbf{H}^{i~-1}\mathbf{y}^{i}$ by thresholding its wavelet coefficients at the scale $j$ according to a threshold vector $\mathbf{t}^{ij}$ defined as 

\begin{equation}
    \mathbf{t}^{ij}_m = \kappa_j \sigma_i \|\mathbf{\Phi}^{j}_{m,:}\mathbf{H}^{i~-1}\|_2
    \label{eq:weights_1}
\end{equation}

\noindent for $j = 1\cdots J$, where $\sigma_i$ is the noise standard deviation in the $i^{th}$ galaxy image and $\kappa_j$ is a scale dependent tuning parameter which can be set as 3 or 4 if the noise is Gaussian. If soft-thresholding is used, this amounts to solving the optimisation problem

\begin{equation}
    \begin{aligned}
        & \underset{\boldsymbol{\alpha}^1,\cdots,\boldsymbol{\alpha}^J}{\text{argmin}}
        & \frac{1}{2}\|\boldsymbol{\Phi}\mathbf{H}^{i~-1}\mathbf{y}^{i}-[\boldsymbol{\alpha}^{1 T},\cdots,\boldsymbol{\alpha}^{J T}]^T\|_2^2+ \sum_{j=1}^J \|\mathbf{t}^{ij}\odot \boldsymbol{\alpha}^j\|_1.
    \end{aligned}
    \label{eq:direct_invert}
\end{equation}  

Such a direct restoration is unsuitable since, as previously mentioned, convolution operators are ill-conditioned, which prevents an accurate inversion. $\mathbf{H}^{i~-1}$, however, can be replaced by $\mathbf{H}^{i T}$, which gives

\begin{equation}
    \mathbf{t}^{ij}_m = \kappa_j \sigma_i \|\mathbf{\Phi}^{j}_{m,:}\mathbf{H}^{i T}\|_2.
    \label{eq:weights_2}
\end{equation}

\noindent The initial weighting can thus be defined as

\begin{equation}
	\mathbf{W}^{(0)}_{:,i} = [\mathbf{t}^{i1 T},\cdots,\mathbf{t}^{iJ T}]^T,
\end{equation}

\noindent for $i = 0\cdots n$. 

This weighting allows one to penalise the wavelet coefficients according to the propagated uncertainty. On the other hand, the reweighting scheme decreases the penalty on the wavelet coefficients that are largely above their expected noise level, which mitigates the bias induced by the $l_1$ norm.

Assuming a white Gaussian noise distribution across the whole field-of-view, the noise standard deviation can be estimated as $\sigma_{est}=1.4826\times\text{MAD}(\mathbf{Y})$ where MAD stands for median absolute deviation and

\begin{equation}
    \text{MAD}((\mathbf{x}_i)_{1\leq i\leq l}) = \text{median}((|\mathbf{x}_i-\text{median}((\mathbf{x}_i)_{1\leq i\leq l})|)_{1\leq i\leq l}).
\end{equation}

In practice, the number of scales, $J$, was set to 3, the tuning parameters in Eq.~\ref{eq:weights_2} were chosen as $\kappa_1=3$, $\kappa_2=3$, and $\kappa_3=4$, and the number of reweightings was set to 1.

%--------------------------------------------------------------------

\section{Low-rank prior}
\label{sec:lowr}

\subsection{Low-rank approximation as a regularisation prior}
Solving Eq.~\ref{eq:rw_l1} is equivalent to deconvolving the detected galaxies independently from one another. A novel way of approaching astronomical image deconvolution is to take advantage of the similarity between galaxy images in a joint restoration process. Indeed, the similar nature of the various galaxy images increases the degeneracy and thus reduces the rank of the matrix $\mathbf{X}$. Therefore, using the prior knowledge that the solution must be of reduced rank can also be used to regularise the inverse problem.

The rank of a matrix can be determined simply by counting the number of non-zero singular values after decomposition ($\mathbf{X} = \mathbf{U}\boldsymbol{\Sigma} \boldsymbol{V}^H$, $^H$ denoting the Hermitian transpose). One may naively assume then that the optimisation problem can be regularised by minimising the rank of the reconstruction but, as with the $l_0$ norm in sparse regularisation, this is a non-convex function and computationally hard to solve. Consequently, the \emph{nuclear norm}, 
    
\begin{equation}
	\|\mathbf{X}\|_* = \sum_{k=1}^{\min(n+1,p)} \sigma_k(\mathbf{X})
\end{equation}

\noindent where $\sigma_k(\mathbf{X})$ denotes the $k^{\text{th}}$ largest singular value of $\mathbf{X}$, is used instead to promote low-rank solutions. 

When combined with Eq.~\ref{eq:l2minpos}, this constraint gives the minimisation problem

\begin{equation}
    \begin{aligned}
        & \underset{\mathbf{X}}{\text{argmin}}
        & \frac{1}{2}\|\mathbf{Y}-\mathcal{H}(\mathbf{X})\|_2^2 + \lambda\|\mathbf{X}\|_*
        & & \text{s.t.}
        & & \mathbf{X} \ge 0
    \end{aligned}
    \label{eq:lowrl2minpos}
\end{equation}   

\noindent where $\lambda$ is a regularisation control parameter.

Low-rank techniques have been applied to exoplanet detection by \citet{gonzalez:16}. See \citet{candes:08} for a more complete introduction to low-rank approximations.

\subsection{Low-rank implementation}

For this work, the assumption is made that a catalogue of similar galaxy images can be approximated by a low-rank representation. Minimisation is implemented via Eq.~\ref{eq:lowrl2minpos} and the threshold, $\lambda$, is calculated as

\begin{equation}
    \lambda = \alpha\sigma_{est}\sqrt{\max(n+1,p)}\rho(\mathcal{H})
\end{equation}

\noindent where $\alpha$ is a threshold factor that was set to 1 for this work. The noise estimate, $\sigma_{est}$, is calculated in the same way as for sparse regularisation.

%--------------------------------------------------------------------

\section{Optimisation}
\label{sec:image_rec}

\subsection{Convex minimisation}

In order to tackle the minimisation problems discussed in the previous sections, a Python code was developed that implements the primal-dual splitting technique described in \citet{condat:13}. Specifically, algorithm 3.1 from  \citet{condat:13} is implemented, neglecting the error terms (as shown in algorithm \ref{alg:condat}), which aims to solve problems of the form

\begin{equation}
    \begin{aligned}
        & \underset{\mathbf{x}}{\text{argmin}}
        & [F(\mathbf{X}) + G(\mathbf{X}) + K(\mathcal{L}(\mathbf{X}))]
    \end{aligned}
\end{equation}

\noindent where $F$ is a convex function with gradient $\nabla F$, $G$ and $H$ are functions with proximity operators that can be solved efficiently, and $\mathcal{L}$ is a linear operator.

\begin{algorithm}
    \caption{Choose the proximal parameters $\tau>0$, $\varsigma>0$, the positive relaxation parameter, $\xi$, and the initial estimate $(\mathbf{X}_0, \mathbf{Y}_0)$. Then iterate, for every $k\geq0$}
    \begin{algorithmic}[1]
        \State $\tilde{\mathbf{X}}_{k+1} = \text{prox}_{\tau G}(\mathbf{X}_k - \tau\nabla F(\mathbf{X}_k) - \tau \mathcal{L}^*(\mathbf{Y}_k))$
        \State $\tilde{\mathbf{Y}}_{k+1} = \mathbf{Y}_k + \varsigma \mathcal{L}(2\tilde{\mathbf{X}}_{k+1} - \mathbf{X}_k) - \varsigma \text{prox}_{K/\varsigma}\Big(\frac{\mathbf{Y}_k}{\varsigma} + \mathcal{L}(2\tilde{\mathbf{X}}_{k+1} - \mathbf{X}_k)\Big)$
        \State $(\mathbf{X}_{k+1}, \mathbf{Y}_{k+1}) := \xi(\tilde{\mathbf{X}}_{k+1}, \tilde{\mathbf{Y}}_{k+1}) + (1 - \xi)(\mathbf{Y}_{k}, \mathbf{Y}_{k})$
    \end{algorithmic}
    \label{alg:condat}
\end{algorithm}

In algorithm \ref{alg:condat}, $\mathbf{X}$ and $\mathbf{Y}$ are the \emph{primal} and \emph{dual} variables respectively. Upon convergence of the algorithm the primal variable will correspond to the final solution (\emph{i.e.} the stack of deconvolved galaxy images). 

For all implementations of this algorithm the primal proximity operator, prox$_{\tau G}$, is the positivity constraint and $\nabla F(\mathbf{X}) = \mathcal{H}^*(\mathcal{H}(\mathbf{X}) - \mathbf{Y})$. $\mathcal{H}^*(\mathbf{Z}) = [\mathbf{H}_0^T\mathbf{Z}_{:,0}, \mathbf{H}_1^T\mathbf{Z}_{:,1}, \dotsc,\mathbf{H}_n^T\mathbf{Z}_{:,n}]$.
   
For sparse regularisation, the dual proximity operator, prox$_{K/\varsigma}$, corresponds to a soft-thresholding with respect to the weights, $\mathbf{W}^{(k)}$ in Eq.~\ref{eq:rw_l1}, and the linear operator, $\mathcal{L}$, corresponds to the starlet transform, $\Phi$. Assuming that the matrices $\mathbf{\Phi}^{j}$ in Eq.~\ref{eq:wav_mat} are circulant, the following inequality holds

\begin{equation}
    \rho(\Phi) \leq \sum_{j=1}^J \|\mathbf{\Phi}^{j}_{0,:}\|_1.
\end{equation}
      
For low-rank regularisation, the dual proximity operator, prox$_{K/\varsigma}$, corresponds to a hard-thresholding of the singular values by the threshold, $\lambda$. The linear operator, $\mathcal{L}$, corresponds to the identity operator, $I_n$, hence $\rho(\mathcal{L})=1$. 

The proximal parameters were set to  $\varsigma = \tau = 0.5$ and the relaxation parameter was set to $\xi = 0.8$ for all implementations.

Convergence was obtained when the change in the cost function was less than $0.0001$ between iterations.

%--------------------------------------------------------------------

\section{Data}
\label{sec:data}
 
\subsection{Galaxy images}
    
The 10\, 000 galaxy images used for the work presented in this paper were obtained from data provided for the GREAT3 challenge \citep{mandelbaum:14}\footnote{http://great3challenge.info/}. Specifically,  the single epoch real space-based galaxy images with a constant PSF. GREAT3  was a galaxy shape measurement challenge with the aim of improving the quality of weak gravitational lensing analysis. The challenge used COSMOS data \citep{koekemoer:07, scoville:07b, scoville:07a} obtained using the Advanced Camera for Survey (ACS) on the Hubble Space Telescope (HST) and processed with the GalSim software package \citep{rowe:15}.

Each galaxy image has a pixel scale of 0.05 arcsec, which is twice the resolution of Euclid (0.1 arcsec pixel scale). This resolution was used to avoid the aliasing issues that will have to be taken into account for the real undersampled Euclid images \citep{cropper:13}. The treatment of these issues is left for future work as the focus of this paper is testing the performance of the deconvolution priors.

Each galaxy is centred within a $96\times96$ pixel postage stamp, but the images were cropped to $41\times41$ to facilitate the use of this data. The pixel flux of the objects range between 0.37 and 814.8, with a median value of 3.7.

These images are well suited to studying Euclid-like observations as the effects of the ACS PSF can be neglected for the purposes of this work, they contain a very small and manageable amount of intrinsic noise, and they are derived from high resolution space-based data.

\subsection{Noise removal}

The intrinsic noise is ``removed'' from each galaxy image, $\mathbf{x}$, by applying soft-thresholding to each image pixel, $\mathbf{x}_i$,

\begin{equation}
	ST_{\lambda}(\mathbf{x}_i) = 
	\begin{cases} 
        \mathbf{x}_i - \lambda\text{sign}(\mathbf{x}_i) & \text{if}\ |\mathbf{x}_i| \geq \lambda \\ 
		0 & \text{otherwise} \\
	\end{cases},
	\label{eq:soft}
\end{equation}
	
\noindent where $\lambda = (1 - w_i) \times \kappa \times \sigma_{est} $, $w_i$ are pixel weights calculated based on local pixel value correlation, $\kappa=4$ and $\sigma_{est}$ is an estimate of the noise.
    
The noise is estimated by taking  the median absolute deviation (MAD) of the starlet transformed image, 

\begin{equation}
    \sigma_{est} = \frac{1.4826 \times MAD(\boldsymbol{\Phi}^{1}\mathbf{x})}{\|\boldsymbol{\Phi}^{1}_{0,:}\|_2}.
\end{equation}

\subsection{Euclid PSFs}

The space variant PSFs used are those described in \citet{kuntzer:16}. In total there are 600 unique PSFs corresponding to different positions across the 4 CCD chips of the Euclid VIS instrument \citep{cropper:12}. The PSFs were simulated using the VIS pipeline and each one has a resolution 12 times that of Euclid. 

\begin{figure}
    \centering
    \includegraphics[width=9cm]{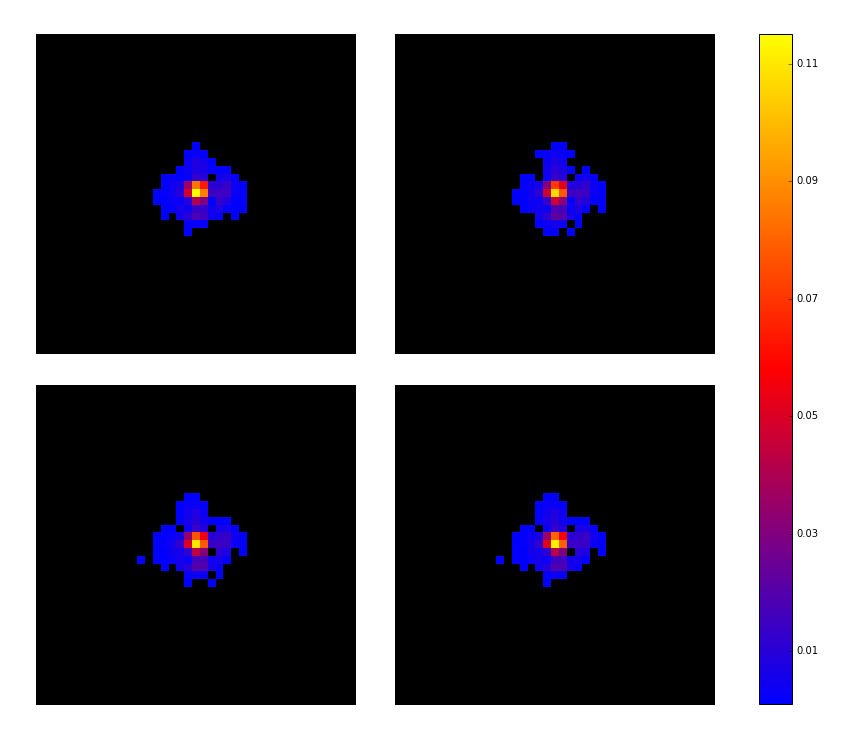}
    \caption{Examples of 4 space variant Euclid-like PSFs.}
    \label{fig:psf_plots}
\end{figure}

The PSFs are down-sampled by a factor of 6 to match the resolution of the galaxy images. Some examples of the downsampled Euclid-like PSFs are shown in Fig.~\ref{fig:psf_plots}. These images demonstrate the anisotropy and diversity of the PSFs used.

\subsection{PSF convolution and Gaussian noise}
 
To produce a stack of Euclid-like observations, $\mathbf{Y}$, each image in the stack of cleaned images, $\mathbf{X}$, is normalised such that the pixel values sum to 1.0, it is then convolved with a random Euclid-like PSF (from the sample of 600) and finally Gaussian noise is added. Gaussian additive noise does not encompass all the predicted sources of noise for Euclid VIS images, however this is a reasonable approximation given that the dominant source of noise expected is readout noise \citep{cropper:13}. 

For this work 5 values of $\sigma$ (the noise standard deviation) were chosen such that the observed galaxies have fixed SNR values of 1.0, 2.0, 3.0, 5.0 and 10.0 (\emph{i.e.} 5 samples of 10\, 000 galaxy images each of which has a fixed SNR for every galaxy).
 
%--------------------------------------------------------------------

\section{Application to data}
\label{sec:data_app}

 \subsection{Quality metrics}

Two metrics are implemented to test the quality of the stack of galaxy images after deconvolution, $\hat{\mathbf{X}}$. The first is the median pixel error, which gives a measure of how similar the deconvolved images are to the original images. The pixel error is given by

\begin{equation}
    P_{err} = \text{median}\left(\frac{\|\mathbf{x}^i - \hat{\mathbf{x}}^i\|_2^2}{\|\mathbf{x}^i\|_2^2}\right)_{1 \leq i \leq n}
    \label{eq:pix_err}
\end{equation}

\noindent A weighted version of the metric is implemented in appenidx~\ref{sec:weighted_perr}.

The second metric is the median ellipticity error, which gives a measure of how well the galaxy shapes can be estimated from the deconvoled images with respect to the clean images. The ellipticity error is given by

\begin{equation}
    \varepsilon_{err} = \text{median}\left(\| \varepsilon(\mathbf{x}^i) - \varepsilon(\hat{\mathbf{x}}^i) \|_2 \right)_{1 \leq i \leq n}
    \label{eq:e_err}
\end{equation}

\noindent where $\varepsilon = [\varepsilon_1, \varepsilon_2]$ is a measure of the ellipticity (or shape) of the galaxy image. Details on how the ellipticities were calculated are provided in appendix~\ref{sec:ellip}.

The regularisation technique that produces the stack of galaxy images with lower values of $P_{err}$ and $\varepsilon_{err}$ at a given SNR is considered to have the better performance.
    
\subsection{Results}

\begin{figure*}
    \centering
    \includegraphics[width=9cm]{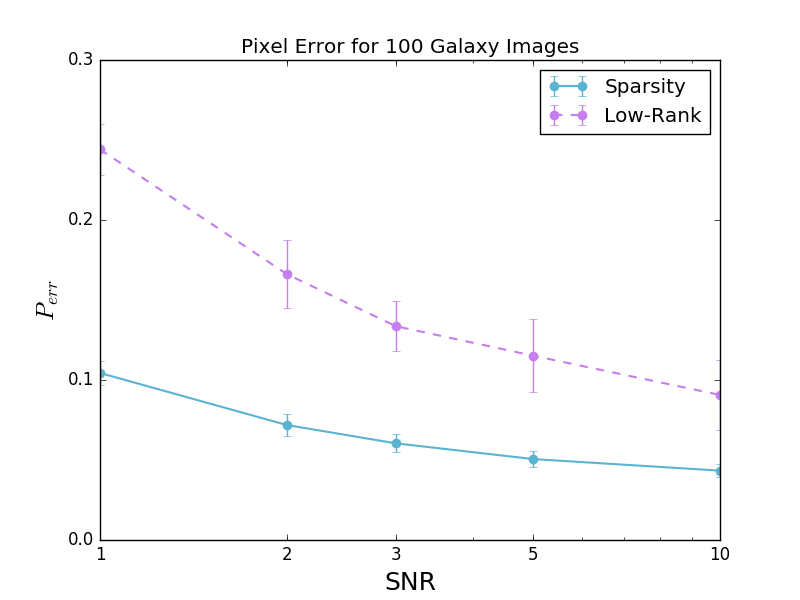}
    \includegraphics[width=9cm]{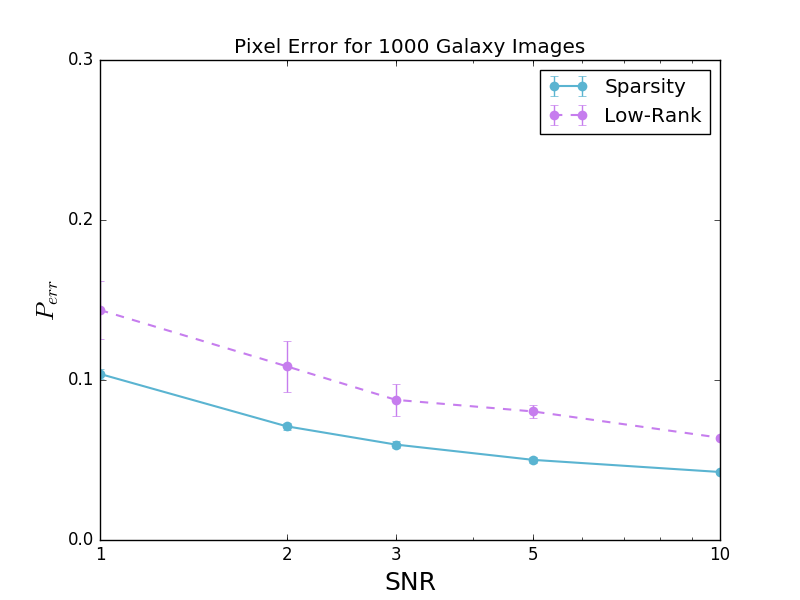}
    \includegraphics[width=9cm]{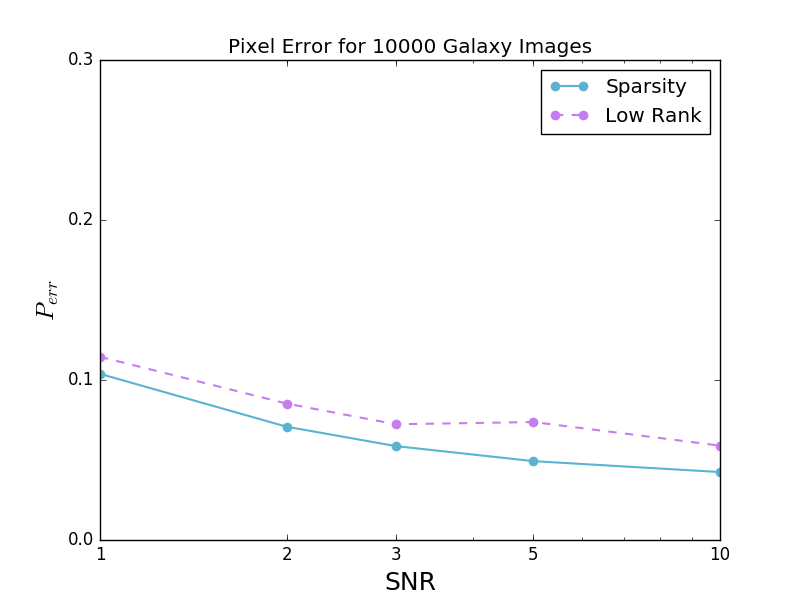}
    \caption{Pixel error as a function of SNR. Mean results from 10 random samples of 100 galaxy images with standard deviation error bars (\emph{top-left panel}). Mean results from 10 random samples of 1000 galaxy images with standard deviation error bars (\emph{top-right panel}). Results for all 10\, 000 galaxy images (\emph{bottom}). Solid blue lines indicate results obtained using sparse regularisation and dashed purple lines indicate results obtained using the low-rank approximation.}
    \label{fig:pixel_curves}
\end{figure*}

\begin{figure*}
    \centering
    \includegraphics[width=9cm]{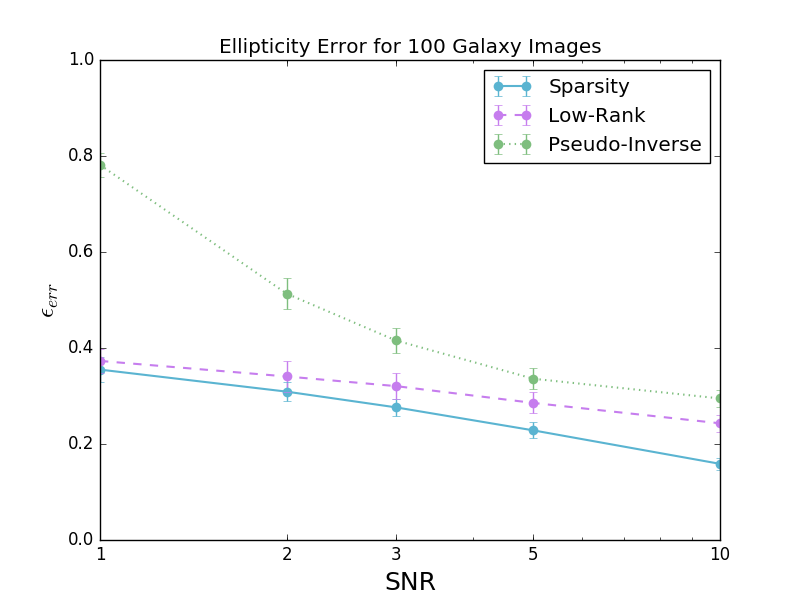}
    \includegraphics[width=9cm]{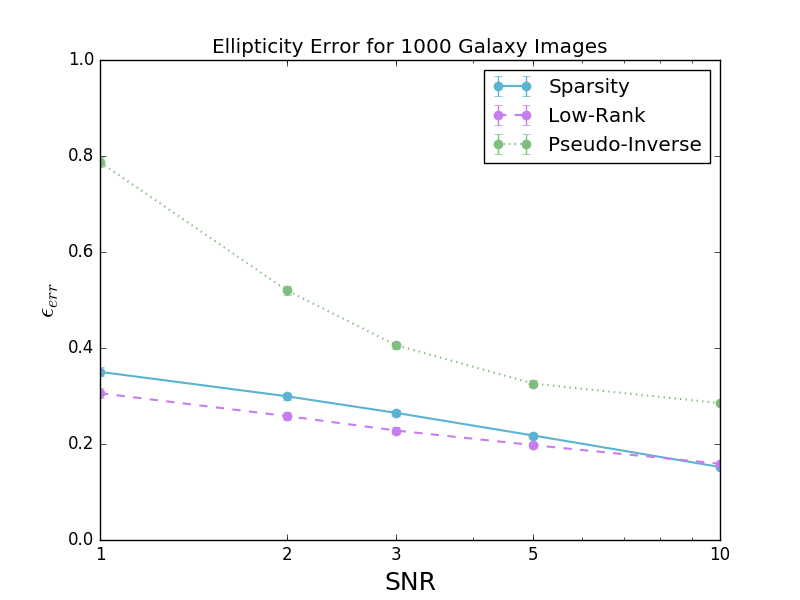}
    \includegraphics[width=9cm]{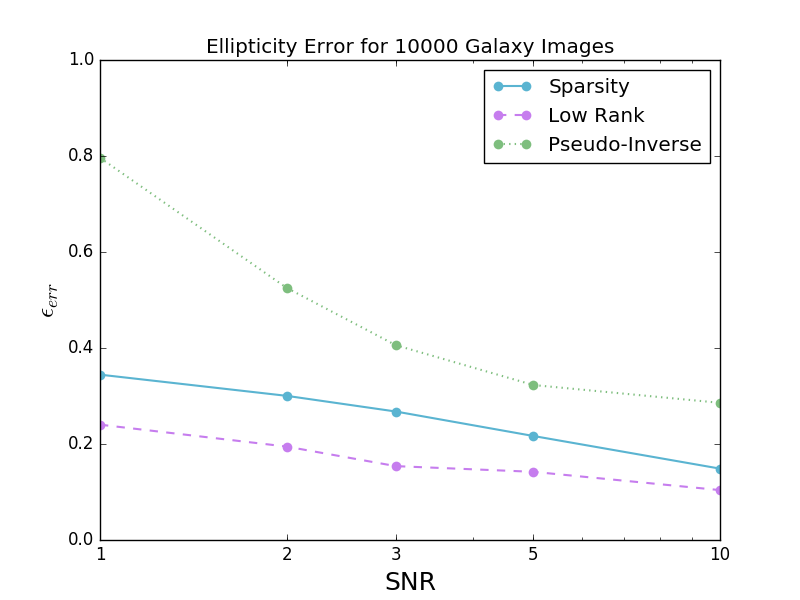}
    \caption{Ellipticity error as a function of SNR. Mean results from 10 random samples of 100 galaxy images with standard deviation error bars (\emph{top-left panel}). Mean results from 10 random samples of 1000 galaxy images with standard deviation error bars (\emph{top-right panel}). Results for all 10\, 000 galaxy images (\emph{bottom}). Solid blue lines indicate results obtained using sparse regularisation, dashed purple lines indicate results obtained using the low-rank approximation and dotted green lines indicate results obtained using a pseudo-inverse deconvolution.}
    \label{fig:ellip_curves}
\end{figure*}

The results of applying the deconvolution code to the data described in Sect.~\ref{sec:data} are shown in Fig.~\ref{fig:pixel_curves} and Fig.~\ref{fig:ellip_curves}. In these figures solid blue lines indicate results obtained using sparse regularisation and dashed purple lines indicate results obtained using the low-rank approximation. In both figures, the top-left and top-right panels show the mean results with standard deviation error-bars as a function of SNR for 10 random samples of 100  and 1000 galaxy images respectively. The bottom panels show the results as a function of SNR for the full sample of 10\, 000 galaxy images. Fig.~\ref{fig:ellip_curves} contains an additional curve in each panel (green dotted lines) that shows the ellipticities obtained from a pseudo-inverse deconvolution (see appendix~\ref{sec:pseudo} for details). Techniques similar to this are commonly implemented in weak lensing analysis to measure galaxy ellipticities.

With regards to the pixel error, sparsity appears to produce better results than the low-rank approach, however the low-rank results improve significantly (by $\sim 10\%$) as the number of galaxy images increases from 100 to 10\, 000. When the full sample is used the difference in the pixel error between the two techniques is around $2\%$.

For the ellipticity error, sparsity performs better when only 100 galaxies are used. With 1000 galaxies the low-rank results show an improvement of a few percent with respect to sparsity and when all 10\, 000 are used low-rank regularisation can provide up to a $10\%$ gain in the ellipticity measurements for low SNR. In all cases the low-rank method performs better than the pseudo-inverse in terms of ellipticity error.

\begin{figure*}[p]
    \centering
    \includegraphics[width=9cm]{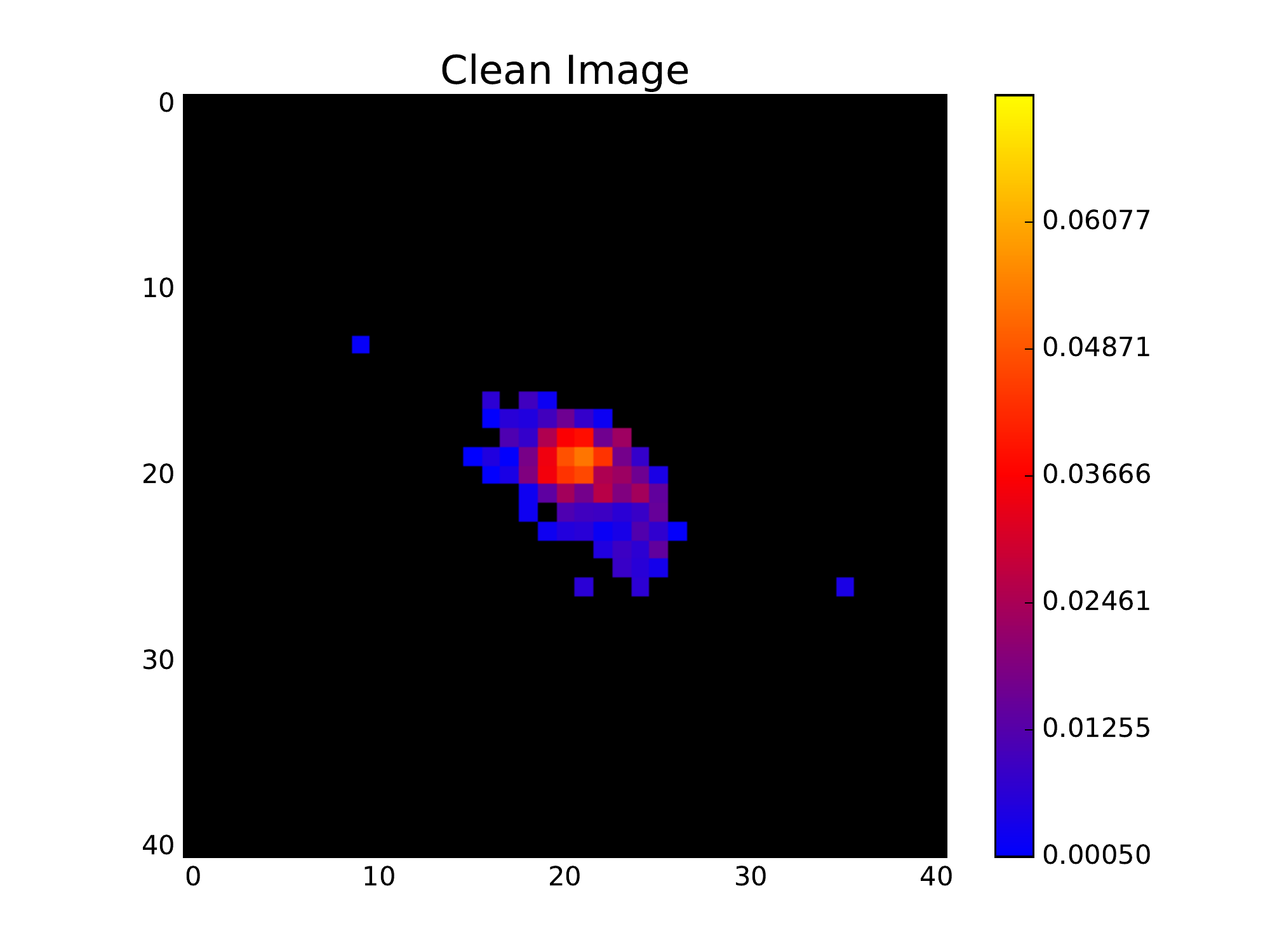}
    \includegraphics[width=18cm]{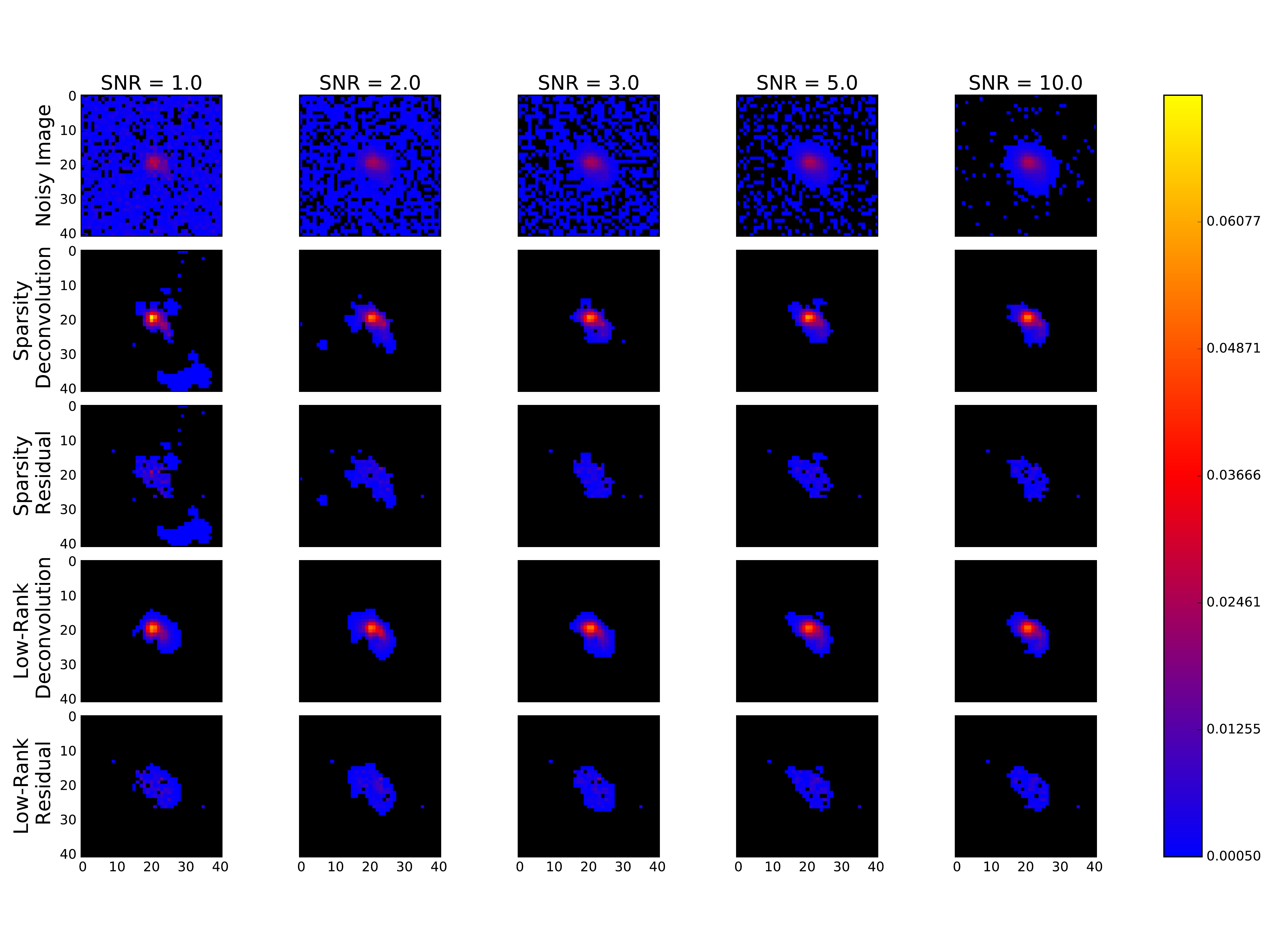}
    \caption{Galaxy 8 of 10\, 000 reference image (\emph{top pannel}), the image convolved with a Euclid-like PSF and various levels of Gaussian noise added (\emph{bottom panel first row}), the image after deconvolution with sparse regularisation (\emph{bottom panel second row}), the residual from sparse deconvolution (\emph{bottom panel third row}), the image after deconvolution with low-rank regularisation (\emph{bottom panel forth row}) and the residual from low-rank deconvolution (\emph{bottom panel fifth row}). Images display the absolute value of the pixels with values less than 0.0005 shown in black.}
    \label{fig:pixel_vals1}
\end{figure*}

\begin{figure*}[p]
    \centering
    \includegraphics[width=9cm]{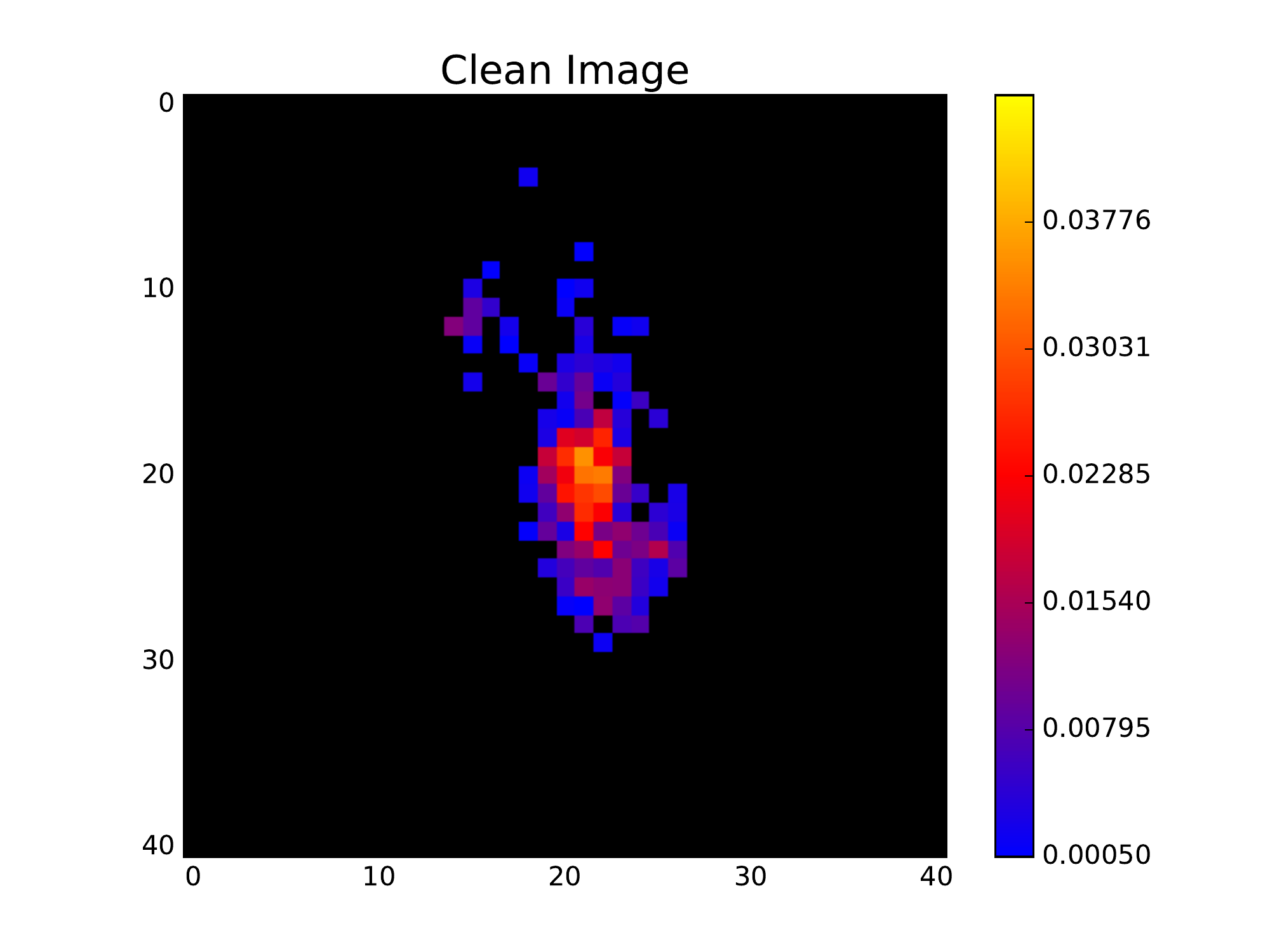}
    \includegraphics[width=18cm]{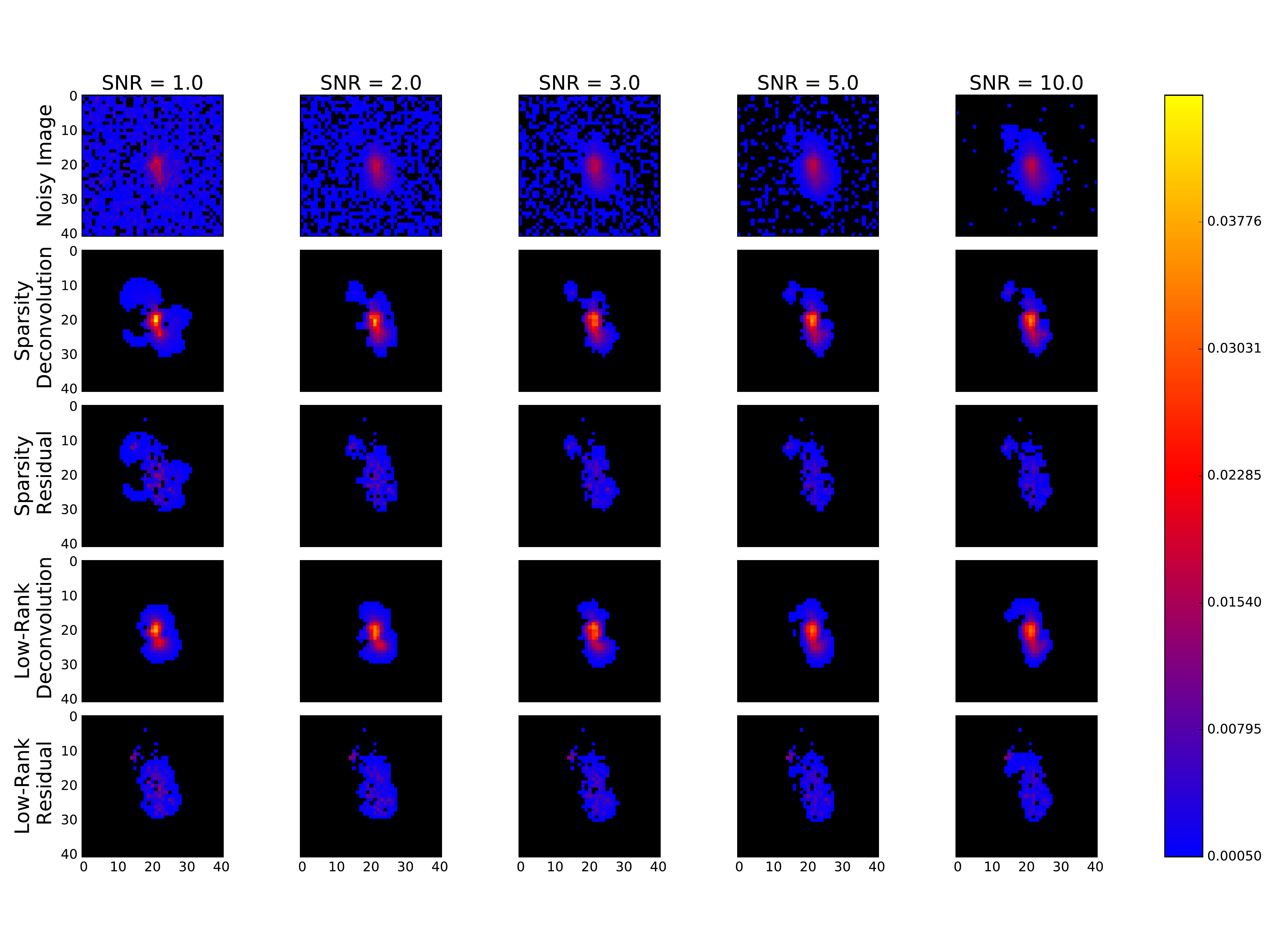}
    \caption{Galaxy 2676 of 10\, 000 reference image (\emph{top pannel}), the image convolved with a Euclid-like PSF and various levels of Gaussian noise added (\emph{bottom panel first row}), the image after deconvolution with sparse regularisation (\emph{bottom panel second row}), the residual from sparse deconvolution (\emph{bottom panel third row}), the image after deconvolution with low-rank regularisation (\emph{bottom panel forth row}) and the residual from low-rank deconvolution (\emph{bottom panel fifth row}). Images display the absolute value of the pixels with values less than 0.0005 shown in black.}
    \label{fig:pixel_vals2}
\end{figure*}

\begin{figure*}[p]
    \centering
    \includegraphics[width=9cm]{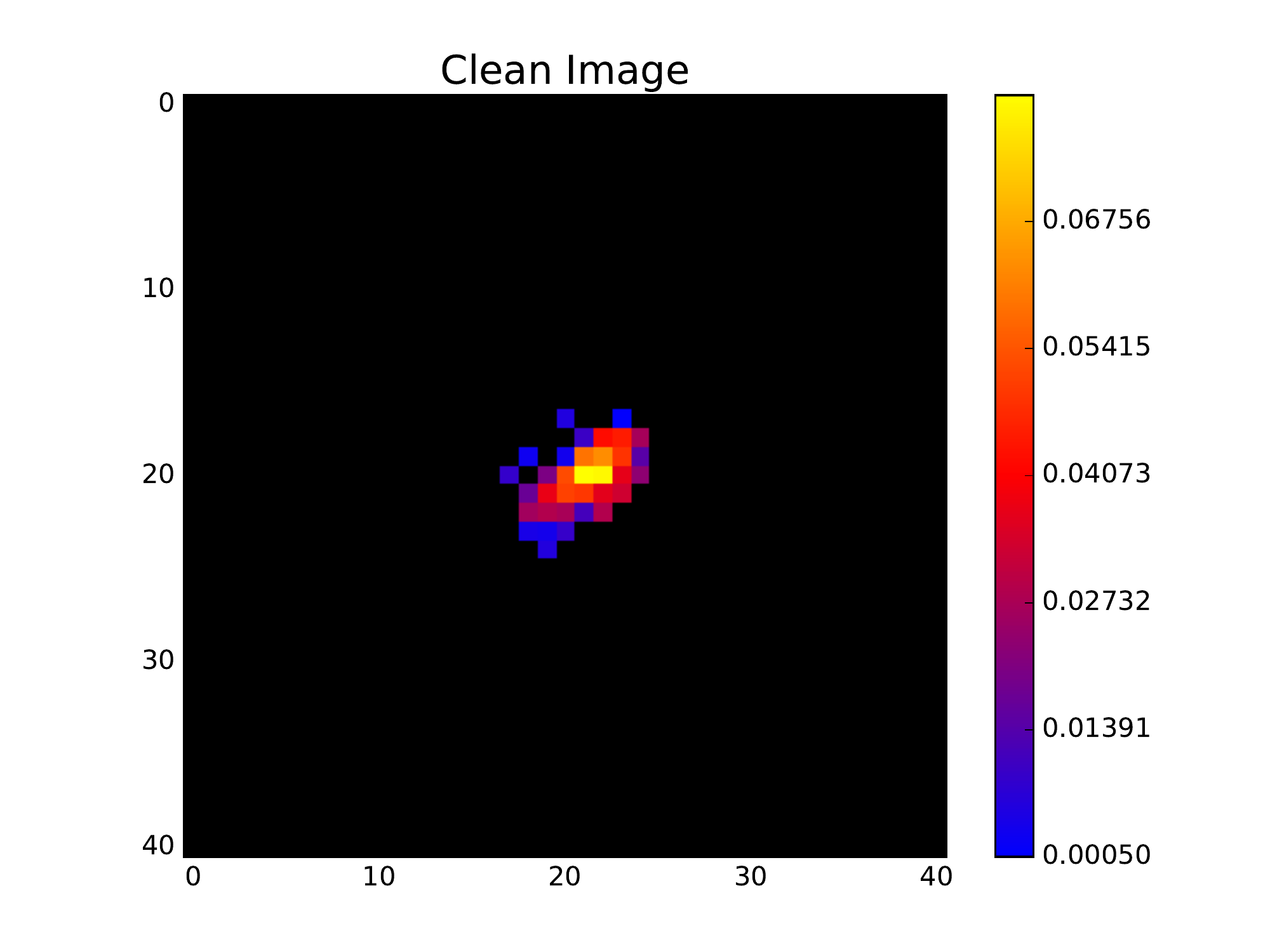}
    \includegraphics[width=18cm]{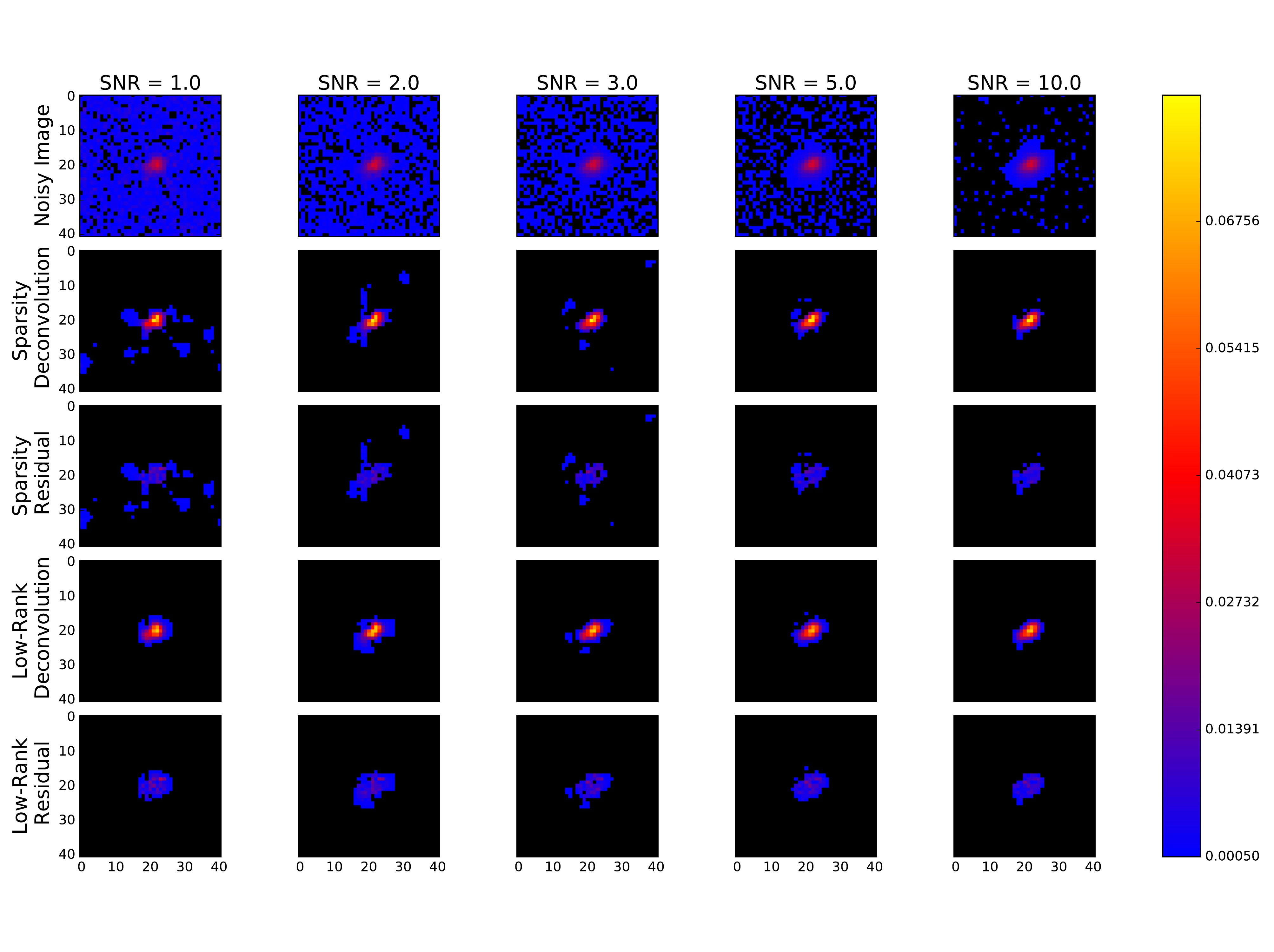}
    \caption{Galaxy 9878 of 10\, 000 reference image (\emph{top pannel}), the image convolved with a Euclid-like PSF and various levels of Gaussian noise added (\emph{bottom panel first row}), the image after deconvolution with sparse regularisation (\emph{bottom panel second row}), the residual from sparse deconvolution (\emph{bottom panel third row}), the image after deconvolution with low-rank regularisation (\emph{bottom panel forth row}) and the residual from low-rank deconvolution (\emph{bottom panel fifth row}). Images display the absolute value of the pixels with values less than 0.0005 shown in black.}
    \label{fig:pixel_vals3}
\end{figure*}

Figures~\ref{fig:pixel_vals1}, \ref{fig:pixel_vals2} and \ref{fig:pixel_vals3} present some examples of individual galaxy images. The top panel in each figure shows the true galaxy images (\emph{i.e.} the original GREAT3 image after intrinsic noise removal). The first row of the lower panel shows the observed galaxy images (\emph{i.e.} convolved with Euclid-like PSF) with various levels of Gaussian noise. The second row shows the images after deconvolution with a sparse prior and the third row shows the corresponding residuals ($\mathbf{x}^i - \hat{\mathbf{x}}^i$). The fourth and fifth rows show the images after deconvolution with a low-rank prior and corresponding residuals respectively. The absolute value of the image pixels are displayed to include negative pixel values and pixels with absolute values below 0.0005 are shown in black for better contrast. In each of these examples both the low-rank and sparsity methods appear to capture the details of the central galaxy pixels, however there is less structure in the tails of the low-rank residuals.

These results demonstrate the potential benefits of exploiting the low-rank matrix approximation for a large sample of galaxy images where many of the images are similar.

\subsection{Convergence speed}

The convergence speed is a factor that was not taken into consideration when comparing the performance of the two regularisation techniques. It should be noted, however, that the low-rank method converged more quickly than the sparsity approach as no re-weighting is required.

%--------------------------------------------------------------------

\section*{Reproducible research}

In the spirit of reproducible research, the space variant deconvolution code has been made freely available on the CosmoStat website\footnote{\url{http://www.cosmostat.org/deconvolution/}}. The noiseless galaxy images have also been provided along with details on how to repeat the experiments performed in this paper.

%--------------------------------------------------------------------

\section{Conclusions}
\label{sec:conc}

A sample of 10\, 000 PSF-free space-based galaxy images were obtained from within the GREAT3 data sets and the intrinsic noise in these images was removed. The images were then convolved with Euclid-like spatially varying PSFs and different levels of Gaussian noise were added to create a series of galaxy images that would be expected from a survey like Euclid.

It has been demonstrated that, using the object oriented deconvolution approach, a new perspective is open for future survey image deconvolution, where images contain many galaxies, which can be assumed to be lying on a given low dimensional manifold. Therefore, a low-rank approximation can be seen as an alternative approach to the most powerful regularising techniques. A deconvolution code that implements both sparsity and a low-rank approximation  was developed. This code was applied to various samples of the data and the two regularisation methods were compared by examining the relative pixel and ellipticity errors in the resulting images as a function of SNR.

The results show that for ten random samples of 100 images sparsity outperforms the low-rank approach in terms of the galaxy images recovered and their corresponding shapes. For ten random samples of 1000 images, the low-rank method performs slightly better with regards to shape recovery. When the full sample of 10\, 000 images is examined, the low-rank method produces significantly lower ellipticity measurement errors particularly for low SNR where the improvement with respect to sparsity is around $10\%$. Also, the degradation in terms of pixel error compared to sparsity is at most $2\%$. This implies that for a sufficiently large sample of images the rank of the matrix can be significantly reduced and used as an effective regularisation prior. This is particularly interesting for projects that require accurate estimates of galaxy shapes.

For future work, it may be interesting to investigate the effects of applying both regularisation techniques simultaneously to see if this can improve the current results. Another interesting study would be to examine the performance when the PSF is not fully known, often the case in real galaxy surveys.

%--------------------------------------------------------------------

\begin{acknowledgements}
    This work is supported by the European Community through the grants PHySIS (contract no. 640174) and DEDALE (contract no. 665044) within the H2020 Framework Program
of the European Commission. The authors wish to thank Yinghao Ge for initial prototyping of the Python code, and Koryo Okumura, Patrick Hudelot and J\'{e}r\^{o}me Amiaux for their work on developing the Euclid-like PSFs. The authors additionally acknowledge the Euclid Collaboration, the European Space Agency and the support of the Centre National d'Etudes Spatiales. Finally, the authors wish to thank the anonymous referee for constructive comments that have improved the quality of the work presented.
\end{acknowledgements}

%--------------------------------------------------------------------

 \bibliographystyle{bibtex/aa}
 \bibliography{bibtex/biblist}
 
%--------------------------------------------------------------------
\begin{appendix}

\section{Weighted pixel error}
\label{sec:weighted_perr}
	
\begin{figure}[ht]
    \centering
    \includegraphics[width=9cm]{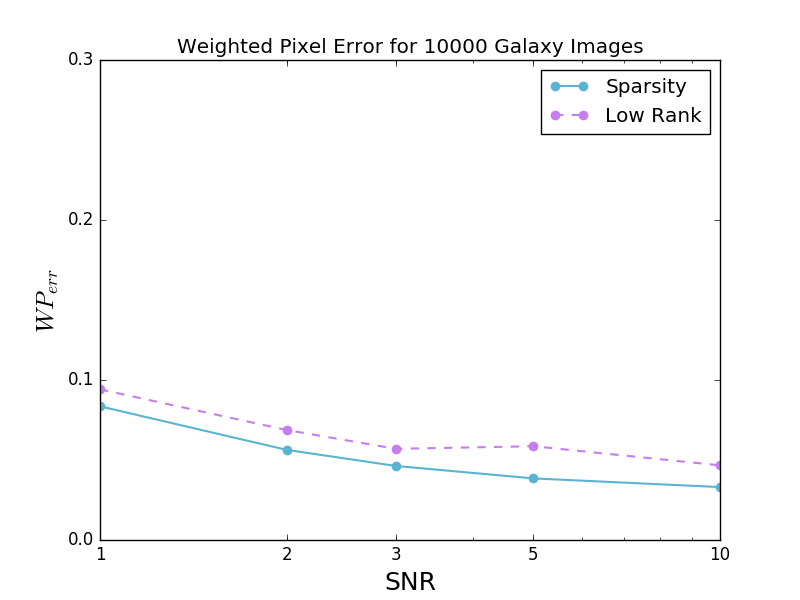}
    \caption{Weighted pixel error as a function of SNR for all 10\, 000 galaxy images. Solid blue lines indicate results obtained using sparse regularisation and dashed purple lines indicate results obtained using the low-rank approximation.}
    \label{fig:pixel_curves_weighted}
\end{figure}

The pixel errors were additionally calculated using the following metric
\\
\begin{equation}
    WP_{err} = \text{median}\left(\frac{\|(\mathbf{x}^i - \hat{\mathbf{x}}^i) 
    \odot \mathbf{w}\|_2^2}{\|\mathbf{x}^i \odot \mathbf{w}\|_2^2}\right)_{1 \leq i \leq n}
    \label{eq:pix_err_weighted}
\end{equation}
\\
\noindent where $\mathbf{w}$ is an isotropic Gaussian kernel with $\sigma=5$. This choice ensures that the central galaxy pixels, which contain most of the information, have higher weights than the tails of the distribution. The results are shown in Fig.~\ref{fig:pixel_curves_weighted}.

The relative performance of the two regularisation techniques is consistent with that shown in Fig.~\ref{fig:pixel_curves} and indicates that sparsity better recovers the central pixel values by a few percent.

\section{Ellipticity measurement}
\label{sec:ellip}
	
Ellipticities were measured following the prescription described in \citet{ngole:16}. The ellipticity components are given by
\\
\begin{equation}
    \varepsilon_1(\mathbf{x}^i) = \frac{<\mathbf{x}^i, \mathbf{U}_4>
    <\mathbf{x}^i, \mathbf{U}_2> - <\mathbf{x}^i, \mathbf{U}_0>^2 + 
    <\mathbf{x}^i, \mathbf{U}_1>^2}{<\mathbf{x}^i, \mathbf{U}_3>
    <\mathbf{x}^i, \mathbf{U}_2> - <\mathbf{x}^i, \mathbf{U}_0>^2 -
    <\mathbf{x}^i, \mathbf{U}_1>^2}
    \label{eq:e1}
\end{equation}
\begin{equation}
    \varepsilon_2(\mathbf{x}^i) = \frac{2\left(<\mathbf{x}^i, \mathbf{U}_5>
    <\mathbf{x}^i, \mathbf{U}_2> - <\mathbf{x}^i, \mathbf{U}_0>
    <\mathbf{x}^i, \mathbf{U}_1>\right)}{<\mathbf{x}^i, \mathbf{U}_3>
    <\mathbf{x}^i, \mathbf{U}_2> - <\mathbf{x}^i, \mathbf{U}_0>^2 -
    <\mathbf{x}^i, \mathbf{U}_1>^2}
    \label{eq:e2}
\end{equation}
\\
\noindent where $<,>$ denotes the inner product, $\mathbf{U}_i$ are shape projection components
\\
\begin{equation}
    \begin{split}
    \mathbf{U}_1 &= (k)_{1 \leq k \leq N_l, 1 \leq l \leq N_c}, \;
    \mathbf{U}_2 = (l)_{1 \leq k \leq N_l, 1 \leq l \leq N_c}, \\
    \mathbf{U}_3 &= (1)_{1 \leq k \leq N_l, 1 \leq l \leq N_c} \;
    \mathbf{U}_4 = (k^2 + l^2)_{1 \leq k \leq N_l, 1 \leq l \leq N_c} \\
    \mathbf{U}_5 &= (k^2 - l^2)_{1 \leq k \leq N_l, 1 \leq l \leq N_c} \;
    \mathbf{U}_6 = (kl)_{1 \leq k \leq N_l, 1 \leq l \leq N_c}
    \end{split}
\end{equation}
\\
\noindent and $N_c$ and $N_l$ correspond to the number of columns and lines in the image $\mathbf{x}^i$ respectively.

It should be noted that equations \ref{eq:e1} and \ref{eq:e2} give identical results to more common implementations \citep[\emph{e.g.}][eq. 12]{cropper:13}.

\section{Pseudo-Inverse Deconvolution}
\label{sec:pseudo}

The pseudo-inverse deconvolution was implemented as
\\
\begin{equation}
    \tilde{\hat{\mathbf{x}}}^i = \frac{\tilde{\mathbf{h}}^{*i} \tilde{\mathbf{y}}^i}
    {\hat{\mathbf{h}}^{*i} \tilde{\mathbf{h}}^i} \tilde{\mathbf{g}}
    \label{eq:pi}
\end{equation}
\\
\noindent where $\tilde{\hat{\mathbf{x}}}^i$, $ \tilde{\mathbf{y}}^i$ and $\tilde{\mathbf{h}}^i$ represent the Fourier transforms of the deconvolved image, the observed image and the PSF respectively. $\tilde{\mathbf{h}}^{*i}$ is the complex conjugate of $\tilde{\mathbf{h}}^i$ and $\tilde{\mathbf{g}}$ is an isotropic Gaussian kernel. For this work a Gaussian kernel with $\sigma = 2$ was used.
    	
\end{appendix}

%--------------------------------------------------------------------

\end{document}